\documentclass[prl,showpacs,superscriptaddress,twocolumn]{revtex4}

\usepackage{amsmath}
\usepackage{amsfonts}
\usepackage{amssymb}

\begin{document}

\title{Giant Casimir effect in fluids in non-equilibrium steady states}

\author{T. R. Kirkpatrick}
\email{tedkirkp@umd.edu}
\affiliation{Institute for Physical Science and Technology, University of Maryland, College Park, Maryland 20877, USA}
\affiliation{Department of Physics, University of Maryland, College Park, Maryland 20877, USA}

\author{J. M. Ortiz de Z\'arate}
\affiliation{Departamento de F\'{\i}sica Aplicada I, Facultad de F\'{\i}sica, Universidad Complutense, 28040 Madrid, Spain}

\author{J. V. Sengers}
\affiliation{Institute for Physical Science and Technology, University of Maryland, College Park, Maryland 20877, USA}

\date{\today}

\begin{abstract}
In this letter we consider the fluctuation induced force exerted between two plates separated by a distance $L$ in a fluid with a temperature gradient. We predict that, for a range of distances $L$, this non-equilibrium force is anomalously large compared to other Casimir forces. The physical reason is that correlations in a non-equilibrium fluid are generally of longer range than other correlations, even than those near an equilibrium critical point. This giant Casimir force is related to a divergent Burnett coefficient that characterizes an Onsager cross effect between the pressure and the temperature gradient. The predicted Casimir force should be detectable with currently available experimental techniques.
\end{abstract}


\maketitle

Fluctuation induced forces are common in nature~\cite{REF1}. The well known prototype of such a force is the Casimir force between conducting plates due to quantum fluctuations of the electromagnetic (EM) field~\cite{REF2}. In this case the overall energy scale is set by Planck's constant, $\hbar$, and the force per unit area, or pressure, is
\begin{equation}
p_{{\text{EM}}}  =  - \frac{{\pi^2 \hbar c}}{{240L^4 }},
\end{equation}
where $L$ is the distance between the plates, $c$ the speed of light, and where the minus sign indicates an attractive force. More recently, non-equilibrium electromagnetic fluctuations when the two plates are at different temperatures, have  been considered~\cite{AntezzaEtAl,Kruger}. Other commonly discussed induced forces involve thermal fluctuations where the energy scale is set by $k_\text{B}T$, where $k_\text{B}$ is Boltzmann's constant and $T$ the temperature~\cite{REF1}. These thermal forces are important when the fluctuations are large and long range. The first instance of this type was noticed by Fisher and de Gennes, who considered finite-size corrections to the free energy in a fluid near a critical point~\cite{REF3}. One then finds a scale-dependent force per unit area, or critical Casimir pressure, $p_{\rm{c}}$, that is given by~\cite{REF4}
\begin{equation}
p_{\rm{c}}  = \frac{{k_{\rm{B}} T}}{{L^3 }}\Theta \left( {L/\xi } \right),
\end{equation}
where $\Theta \left( x \right)$ is a finite-size scaling function with $\xi$ the correlation length. One defines a Casimir amplitude, $\Delta  = \mathop {\lim }\limits_{x \to 0} \Theta \left( x \right)$, which for the Ising-like universality class may vary from $-0.01$ to $+2$ depending on the boundary conditions~\cite{REF5}. Note that at larger $L$, $|p_\text{c}| > |p_\text{EM}|$. That is, the fluctuations that cause $p_\text{c}$ are effectively of longer range than those that cause $p_\text{EM}$. Similar Casimir forces have been predicted to exist generally in equilibrium systems, when long-range correlations are present due to the existence of Goldstein modes. Systems investigated include superfluid helium and liquid crystals~\cite{REF1}.

It has by now been well established that thermal fluctuations in fluids in non-equilibrium steady states are anomalously large and very long range. The most studied case is a quiescent fluid in the presence of a uniform temperature gradient, $\nabla{T}$. Then the non-equilibrium contribution to the temperature fluctuations as a function of the wave number, $k$, is given by~\cite{REF6}
\begin{equation}
\left\langle {\left| {\delta T\left( \mathbf{k} \right)} \right|^2 } \right\rangle _{{\rm{NE}}}  = \frac{{k_{\rm{B}} T}}{{\rho D_T \left( {\nu  + D_T } \right)}}\frac{{\left( {k_\parallel  \nabla T} \right)^2 }}{{k^6 }}.
\end{equation}
Here the temperature gradient is taken to be in the $z$-direction, the plates are located at $z=0,L$, and $k_\parallel$ is the magnitude of the component of the wave vector $\mathbf{k}$ in the direction parallel to the plates. In this equation, $\rho$ is the mass density,  $D_T$ the thermal diffusivity, and $\nu$ the kinematic viscosity. This result for the temperature fluctuations at hydrodynamic length scales was predicted a long time ago~\cite{REF7,REF8} and has been verified accurately by light scattering \cite{REF9,REF10} and by shadow-graph experiments~\cite{REF11}. From Eq. (3) one sees that the intensity of the temperature fluctuations diverges as $k^{-4}$ when $k\to0$. The purpose of this Letter is to show that, as a consequence, these non-equilibrium temperature fluctuations will cause a Casimir effect that is even more significant than the one induced by critical fluctuations, whose intensity only varies as $k^{-2}$~\cite{REF12}. Specifically, for the scale-dependent non-equilibrium fluctuation contribution, $p_{{\rm{NE}}}$, to the pressure we have found
\begin{equation}\begin{split}
\hspace*{-20pt}p_{{\rm{NE}}}  &= \frac{{c_p k_{\rm{B}} T^2 \left( {\gamma  - 1} \right)}}{{96\pi D_T \left( {\nu  + D_T } \right)}}\\ &\times \left[ {1 - \frac{1}{{\alpha c_p }}\left( {\frac{{\partial c_p }}{{\partial T}}} \right)_p  + \frac{1}{{\alpha ^2 }}\left( {\frac{{\partial \alpha }}{{\partial T}}} \right)_p } \right]L\left( {\frac{{\nabla T}}{T}} \right)^2.
\end{split}\end{equation}
Here $c_p$ is the isobaric specific heat capacity, $\gamma$ the ratio of the isobaric and isochoric heat capacities, and $\alpha$ the thermal expansion coefficient. Note that for a fixed temperature gradient, this Casimir force actually grows with increasing $L$. This anomalous behavior is a reflection of the very long spatial correlations in a fluid in a non-equilibrium state. Experimentally, it may be easier to measure the $L$ dependence of the non-equilibrium Casimir pressure by fixing the temperature difference, $\Delta T$, between the two plates and varying the distance $L$. If a uniform temperature gradient is present, then $\nabla T = \Delta T / L$, and for fixed $\Delta T$ the non-equilibrium Casimir effect decreases as $L^{-1}$ for larger $L$. Also note that $p_{{\rm{NE}}}$ can be positive or negative.

To understand the physical origin of Eq. (4), we start with the general observation that a temperature gradient can cause normal stresses or pressures, if nonlinear effects are taken into account. Specifically, we consider a non-linear Onsager cross effect characterized by a non-linear kinetic coefficient that we define as $\kappa _{{\rm{NL}}}$. On these general grounds one expects a non-equilibrium contribution to the pressure that is given by
\begin{equation}
p_{{\rm{NE}}}  = \kappa _{{\rm{NL}}} \left( {\nabla T} \right)^2.
\end{equation}

In standard transport theory, $\kappa _{{\rm{NL}}}$ is referred to as a non-linear Burnett coefficient~\cite{REF13}. It is well known that these transport coefficients, which go beyond ordinary Navier-Stokes transport coefficients, do not exist due to the presence of long-time-tail (LTT) effects~\cite{REF6,REF14}. Indeed, at this order one expects that $\kappa _{{\rm{NL}}}$ has a LTT contribution that diverges linearly with the system size~\cite{REF15,REF16}. To account for such a LTT contribution, one should consider $\kappa _{{\rm{NL}}}$ as the sum of a bare contribution, $\kappa _{{\rm{NL}}}^{\left( 0 \right)}$, associated with short-range correlations, and a divergent contribution $\sim L$ due to long-range correlations with a coefficient $\kappa _{{\rm{NL}}}^{\left( 1 \right)}$. Substituting $\kappa _{{\rm{NL}}}  \simeq \kappa _{{\rm{NL}}}^{\left( 0 \right)}$ into Eq. (5) yields a pressure effect associated with short-range correlations. Since the ratio $\kappa _{{\rm{NL}}}^{\left( 0 \right)} /\kappa _{{\rm{NL}}}^{\left( 1 \right)} L$ will be of the order of $\sigma /L$, where $\sigma$ is a typical intermolecular distance, this contribution is small and can be neglected. Also such an effect is one of several at molecular scales, including accommodation of the velocity and kinetic energy of the molecules with the wall, which do not satisfy universal laws and should not be characterized as Casimir effects~\cite{REF4}. However, substitution of $\kappa_{{\rm{NL}}}  \simeq \kappa _{{\rm{NL}}}^{\left( 1 \right)} L$ into Eq.~(5) yields a genuine non-equilibrium Casimir pressure due to long-range correlations:
\begin{equation}
p_{{\rm{NE}}}  = \kappa _{{\rm{NL}}}^{\left( 1 \right)} L\left( {\nabla T} \right)^2.
\end{equation}
Hence, Eq.~(4) represents a calculation of the LTT, or divergent, part of $\kappa _{{\rm{NL}}}$. It thus follows that a measurement of the non-equilibrium Casimir pressure, $p_{{\rm{NE}}}$, will not only verify the existence of a novel type of Casimir effect, but would also be a direct measure of a divergent Burnett coefficient.

Our main result, Eq.~(4), can be derived by two distinct approaches. The first approach uses a general statistical-mechanical method~\cite{REF17} to express the normal stress in terms of a non-equilibrium time-correlation function:
\begin{equation}
p_{{\rm{NE}}}  =  - \frac{1}{{k_{\rm{B}} T^2 }}\nabla T\int\limits_0^\infty  {dt\left\langle {{J}_l \left( t \right)~ {J}_\varepsilon  \left( 0 \right)} \right\rangle } _{{\rm{LE}}}.
\end{equation}
Here ${J}_l$ is a microscopic longitudinal stress current, ${J}_\varepsilon$ a microscopic energy current, and the angular brackets denote an average over an $N$-particle local equilibrium (LE) distribution function. This time correlation function can then be evaluated in what amounts to a mode-coupling approximation by using techniques developed previously~\cite{REF18,REF19}. The chief difference with equilibrium mode-coupling theory is that in this case one must use hydrodynamic modes appropriate for the non-equilibrium steady state considered here. It also means that the modes are constructed for a system finite in the $z$-direction. To obtain an explicit analytic result we assume stress-free boundary conditions for the fluctuations. The net result is Eq.~(4). One sees explicitly that this equation is proportional to the mode-coupling amplitude for the heat conductivity times the one for the longitudinal, \emph{i.e.} bulk, viscosity~\cite{REF20}. That is, the non-equilibrium pressure is an Onsager cross effect with a divergent $\sim L$ Burnett kinetic coefficient.

The second method to derive Eq.~(4) starts from a non-linear fluctuating pressure, relating it to the temperature fluctuations, and then evaluating the temperature fluctuations by, for example, the method of non-equilibrium fluctuating hydrodynamics~\cite{REF21}. It is convenient to consider the pressure as a function of a fluctuating energy density $e + \delta e$ and a fluctuating number density $n+\delta{n}$ and write the pressure in terms of the mean values, $e,n$, and their fluctuations $\delta e, \delta n$:
\begin{equation}
p\left( {e + \delta e,n + \delta n} \right) = p(e,n) + \delta p.
\end{equation}
Taylor expanding this pressure and averaging the result leads to a fluctuation renormalization of $p(e,n)$. The fluctuations that lead to Eq.~(3) arise from a combination of entropy and transverse velocity fluctuations with vanishing linear pressure fluctuations~\cite{REF6,REF7,REF21}. This implies that to second order in the fluctuations the non-equilibrium pressure renormalization is given by
\begin{equation}\begin{split}
p_{{\rm{NE}}}  = \frac{{\left( {n\alpha } \right)^2 }}{2}\left[ \left( {\frac{{\partial ^2 p}}{{\partial n^2 }}} \right)_e  - 2w\left( {\frac{{\partial ^2 p}}{{\partial e\partial n}}} \right)\right.& \\ \left. + w^2 \left( {\frac{{\partial ^2 p}}{{\partial e^2 }}} \right)_n  \right]& \left\langle {\left( {\delta T} \right)^2 } \right\rangle _{{\rm{NE}}}
\end{split}\end{equation}
with
\begin{equation}
w = \left( {\frac{{\partial p}}{{\partial n}}} \right)_e /\left( {\frac{{\partial p}}{{\partial e}}} \right)_n.
\end{equation}
In deriving Eq.~(9) we have used that vanishing linear pressure fluctuations implies $\delta e=-w \delta n$ and $\delta n = -n \alpha \delta T$. To evaluate the temperature fluctuations induced by the presence of a temperature gradient one should notice that the $k^{-4}$  variation of the temperature fluctuations as function of the wave number in Eq.~(3) causes large finite-size effects when the fluid layer is bounded by plates separated by a finite distance $L$. These finite-size effects have been evaluated both for stress-free boundary conditions~\cite{REF22} and for no-slip boundary conditions~\cite{REF23}. If we assume stress-free boundary conditions, for which an explicit analytic expression is available~\cite{REF22}, we obtain
\begin{equation}
\begin{split}
\left\langle {\left( {\delta T} \right)^2 } \right\rangle _{{\rm{NE}}}&= \frac{1}{V}\int {d\mathbf{x}\left\langle {\delta T\left( \mathbf{x} \right)^2 } \right\rangle _{{\rm{NE}}} }
\\& = \frac{{k_{\rm{B}} T}}{{48\pi \rho D_T \left( {\nu  + D_T } \right)}}L\left( {\nabla T} \right)^2.
\end{split}
\end{equation}
Upon substituting Eq.~(11) into Eq.~(9) and using some thermodynamic manipulations we find Eq.~(4) again.

To estimate the non-equilibrium Casimir pressure, $p_{{\rm{NE}}}$, implied by Eq.~(4), we consider as an example liquid toluene for which accurate light-scattering measurements of the non-equilibrium temperature fluctuations are available~\cite{REF9,REF10}. Using information for the thermodynamic and transport properties~\cite{REF10,REF24}, we have calculated from Eq.~(4) values of $p_{{\rm{NE}}}$ in the case of liquid toluene with $\Delta T = 10$~K at an average temperature of 25$^\circ$C. In Table~I we compare these values of $p_{{\rm{NE}}}$ with values for the electromagnetic Casimir pressure, $p_{{\rm{EM}}}$, calculated from Eq.~(1), and the critical Casimir pressure, $p_{\rm{c}}  = k_{\rm{B}} T\Delta /L^3$ with a Casimir amplitude $\Delta = -0.15$ corresponding to periodic boundary conditions~\cite{REF25}, which are conceptually closest to the stress-free boundary conditions adopted in the present paper for the non-equilibrium temperature fluctuations. It is seen that at $L = 1~\mu$m $p_{{\rm{NE}}}$ is one order of magnitude larger than $|p_{{\rm{c}}}|$, and at $L = 1$~mm is seven orders of magnitude larger than $|p_{{\rm{c}}}|$. In terms of a non-equilibrium Casimir force per unit area, $F_{{\rm{NE}}} /A$, with $A = 1$~mm$^2$, a typical value quoted in the literature~\cite{REF4}, it means that at $L=1~\mu$m $F_{{\rm{NE}}}  \simeq 1.5 \times 10^{ - 8} \;{\rm{N}}$, and at $L = 1$~mm  $F_{{\rm{NE}}}  \simeq 1.5 \times 10^{ - 11} \;{\rm{N}}$. With modern experimental resolutions approaching femto-Newtons~\cite{REF4,REF26,REF27}, it may even become possible to measure a non-equilibrium Casimir effect at a distance $L = 1$~mm where the more traditional Casimir effects are unobservable. Another non-equilibrium Casimir effect can be expected in liquid mixtures where a temperature gradient induces long-range concentration fluctuations through the Soret effect~\cite{REF10,REF28,REF29,REF30,REF31}

\begin{table}
\caption{Estimated Casimir pressures}
\begin{tabular*}{\columnwidth}{c@{\extracolsep{\fill}}ccc}
\toprule
$L$&$p_\text{EM}$\footnote{Eq.~(1)}&$p_\text{c}$\footnote{$p_\text{c}=-0.15 k_\text{B}T/L^3$}&$p_\text{NE}$\footnote{Toluene with $\Delta{T}=10$~K} \\
\colrule
$10^{-6}$~m& $-1.3\times10^{-3}$~Pa  & $-1.2\times10^{-3}$~Pa  &$3.0\times10^{-2}$~Pa\\
$10^{-3}$~m& $-1.3\times10^{-15}$~Pa & $-1.2\times10^{-12}$~Pa &$3.0\times10^{-5}$~Pa\\
\botrule
\end{tabular*}
\end{table}

In deriving expression~(4) for the non-equilibrium Casimir pressure, we have not included any possible effect of gravity on the non-equilibrium fluctuations. Gravity does not affect the intensity of the temperature fluctuations at small and large wave numbers. However, gravity suppresses temperature fluctuations at intermediate wave numbers in liquid layers with a negative Rayleigh number and enhances temperature fluctuations in liquid layers with a positive Rayleigh number~\cite{REF32,REF33}.
This means that in liquid layers heated from above the non-equilibrium Casimir pressure will be somewhat smaller and in liquid layers heated from below somewhat larger than that deduced from Eq.~(4), and diverging when the critical value of the Rayleigh number associated with the onset of convection is approached.  In developing experimental techniques for measuring the non-equilibrium Casimir effect, it will be highly desirable to be able to measure not only the magnitude of the proposed non-equilibrium Casimir effect, but also its dependence on the distance $L$, since both quantities are qualitatively different from those of the more traditional Casimir effects.

We note that Najafi and Golestanian~\cite{Golestanian} have considered a non-equilibrium Casimir effect due to inhomogeneous noise correlations in a medium that is otherwise in local equilibrium. However, it is known that non-equilibrium fluctuations arising from inhomogeneous noise correlations are considerably less significant than those arising from hydrodynamic couplings induced in the fluid by a temperature gradient~\cite{miStatis}.

We thank M.L. Huber and E.W. Lemmon at the US National Institute of Standards and Technology for providing us with the relevant thermophysical-property information for saturated liquid toluene and C. V\'elez at the Universidad Complutense for assisting us with obtaining quantitative estimates for $p_{{\rm{NE}}}$. The research was supported by the US National Science Foundation under Grant No. DMR-09-01907.

\end{document}